\documentclass[runningheads,a4paper]{llncs}

\usepackage{amssymb}
\usepackage{graphicx}
\usepackage{mathrsfs}
\usepackage{amsmath}

\usepackage{url}
\usepackage{epstopdf}
\urldef{\mailsa}\path|stys@jcu.cz|
\newcommand{\keywords}[1]{\par\addvspace\baselineskip
\noindent\keywordname\enspace\ignorespaces#1}

\begin{document}

\mainmatter  

\title{Measurement in biological systems from the self-organisation point of view}

\titlerunning{Measurement in biological systems}

%
%
\author{Dalibor \v{S}tys \and Jan Urban \and Renata Rycht\'{a}rikov\'{a} \and Anna Zhyrova \and Petr C\'{i}sa\v{r}}
\authorrunning{Dalibor \v{S}tys et al.}

\institute{Institute of Complex Systems, Faculty of Fisheries and Protection of Waters\\
University of South Bohemia, Z\'{a}mek 136, 373 33 Nov\'{e} Hrady, Czech Republic
\mailsa\\
\url{http://www.frov.jcu.cz/cs/ustav-komplexnich-systemu-uks}}

%
%

\toctitle{Lecture Notes in Bioinformatics}
\tocauthor{Dalibor \v{S}tys, et al.}
\maketitle

\begin{abstract}
Measurement in biological systems became a subject of concern as a consequence of numerous reports on limited reproducibility of experimental results. To reveal origins of this inconsistency, we have examined general features of biological systems as dynamical systems far from not only their chemical equilibrium, but, in most cases, also of their Lyapunov stable states. Thus, in biological experiments, we do not observe states, but distinct trajectories followed by the examined organism. If one of the possible sequences is selected, a minute sub-section of the whole problem is obtained -- sometimes in a seemingly highly reproducible manner. But the state of the organism is known only if a complete set of possible trajectories is known. And this is often practically impossible. Therefore, we propose a different framework for reporting and analysis of biological experiments, reflecting the view of non-linear mathematics. This view should be used to avoid overoptimistic results, which have to be consequently retracted or largely complemented.  An increase of specification of experimental procedures is the way for better understanding of the scope of paths, which the biological system may be evolving. And it is hidden in the evolution of experimental protocols. Our system bioWes is a tool for objectivization of this knowledge.

\end{abstract}

\section{Introduction}
\label{sec:introduction}

Measurement in biological systems became a subject of concern as a consequence of numerous reports on limited reproducibility of experimental results \cite{Prinz2011,Begley2012}. By detailed examination, it was often found that many of the results are not exactly fake, but represent a selection from actually obtained results. In the same time, articles are accompanied by statistical analysis, which seemingly confirms the normal, Gaussian, distribution of results. 

To reveal this inconsistency, we have discussed the main features of biological systems from the mathematical point of view. Biological systems are dynamical systems maintained far from not only their equilibrium, but in most cases also of recurrent, Lyapunov stable states \cite{Lyapunov1892}. In other words, in biological experiments, we do not observe states, but distinct trajectories followed by the examined organism. These trajectories are characterized by a sequence of distinct spatial structures of the organism, i.e., a sequence of cell states. 

There are two points of view from which this problem must be approached: (1) the properties of the experimental system itself and (2) technical potential of the measurement. 

\section{Technical limits of the information content of the measurement} 
\label{sec:system_theory}

A new system theory was introduced by Pavel \v{Z}ampa \cite{Zampa2004}. The main differences of the \v{Z}ampa\'{ }s systems theory from other system theories are (1) inclusions of the input and output into the system description, (2) a definition of the system attribute as a distinct concept from the system variable, and (3) an introduction of the system time as a time of measurement. From that, the concept of the complete immediate cause as the list of values of all system attributes at all time instants, preceding the examined system time necessary for its description, naturally arises. Here we summarise selected parts of \v{Z}ampa\'{ }s system theory needed for discussion in this article. 

The adequate model of the time, which we call a real time,  is a variable  $t$,

\begin{equation}
t\in T,
\label{eq:real_time}
\end{equation}
whose definition set is a non-empty set $T$ of all time events   

\begin{equation}
t_k, \mbox{ where } k\in \{0,1,2,...,F\}.
\label{eq:index_set_of_time}
\end{equation}
If there exist a relation $t_i<t_j$ for each two time instants $t_i, t_j\in T$,  we say that the time instant $t_i$ precedes the time instant $t_j$. 

We shall denominate the studied system attributes (abstract variables) -- such as a coordinate of position, coordinate of speed, position of a switch, verity of a statement -- by symbols $a_i$, where $i=1,2,...,n$. The set of all abstract variables will be denominated by symbol $A$, and thus it holds:

\begin{equation}
A=\{a_i\mid i\in I\},
\label{eq:attributes_set}
\end{equation}
where $I$ is an appropriate nonempty index set. Adequate model of the i-th attribute $a_i, i\in I$ is an abstract variable of the i-th attribute $ v_i $
  
\begin{equation}
v_i\in V_i, \mbox{ for } i\in I,
\label{eq:definition_of_variable}
\end{equation}
whose definition set is a nonempty set $V_i$, where $i\in I$, with elements, which we shall call \textsl{values of the i-th attribute}. The introduction of the variable $t\in T$ and the set of variables $v_i\in V_i$ for $i\in I$, formalised the notion of time and other attributes of the system. We may now introduce \textsl{a system variable} $v$ defined by relations

\begin{equation}
v=(v_1,v_2,...,v_n),
\label{eq:definition_of_system_variable}
\end{equation}

\begin{equation}
v\in V,
\label{eq:system_variable_definition_set}
\end{equation}
and 

\begin{equation}
V=V_1\times V_2\times ... \times V_n
\label{eq:system_variable_in_time}
\end{equation}

Thus, the system variable is an ordered set of $n$ variables of system attributes. Then, the ordered set
 
\begin{equation}
(T,V)
\label{eq:definition_of_the_system_mathematical_base}
\end{equation}
is the basis of the mathematisation of the problem of the definition of the system trajectory as a mapping. The trajectory of an abstract system corresponds to the  mapping $z$

\begin{equation}
z:T\times I\rightarrow \bigcup_{i\in I}V_i\ \mbox{ such that }
\ z(t,i)\in V_i, i\in I
\label{eq:system_trajectory_definition}
\end{equation}
If we denote the set of all system trajectories as $\Omega$, we may write 

\begin{equation}
\Omega = \{z\mid z:T\times I\rightarrow \bigcup_{i\in I}V_i\
\mbox{ such that }\ z(t,i)\in V_i, i\in I\}.
\label{eq:definition_of_the_set_of_trajectories}
\end{equation}

The system event is marked $B$ and defined as a sub-set of the set $\Omega$ of all system trajectories:

\begin{equation}
B\subset \Omega.
\label{eq:definition_of_system_event}
\end{equation}
We usually demarcate the system event as a set of trajectories $z$ having a certain property $V(z)$, which we describe as 

\begin{equation}
 B = \{z\mid V(z)\}.
 \label{eq:property_of_system_event}
\end{equation} 
 Then, we may define an abstract deterministic system ${\mathscr{D}}$ as

\begin{equation}
{\mathscr{D}}=(T,V,z).
\label{eq:definition_of_deterministic_system}
\end{equation}

For technical as well as internal mechanism reasons, neither of the measurable systems is truly deterministic. By introduction of probabilistic instead of deterministic mapping $P(B)$, which is defined on the potency $\cal{B}$ of the set $ \Omega $ of all trajectories $B\in\cal{B}$ we may define stochastic (abstract) system $\mathscr{S} $ as

\begin{equation}
{\mathscr{S}}=(T,V,P).
\label{eq:definition_of_stochastic_abstract_system}
\end{equation}

\begin{figure*}
\centering
\includegraphics[width=0.53\textwidth]{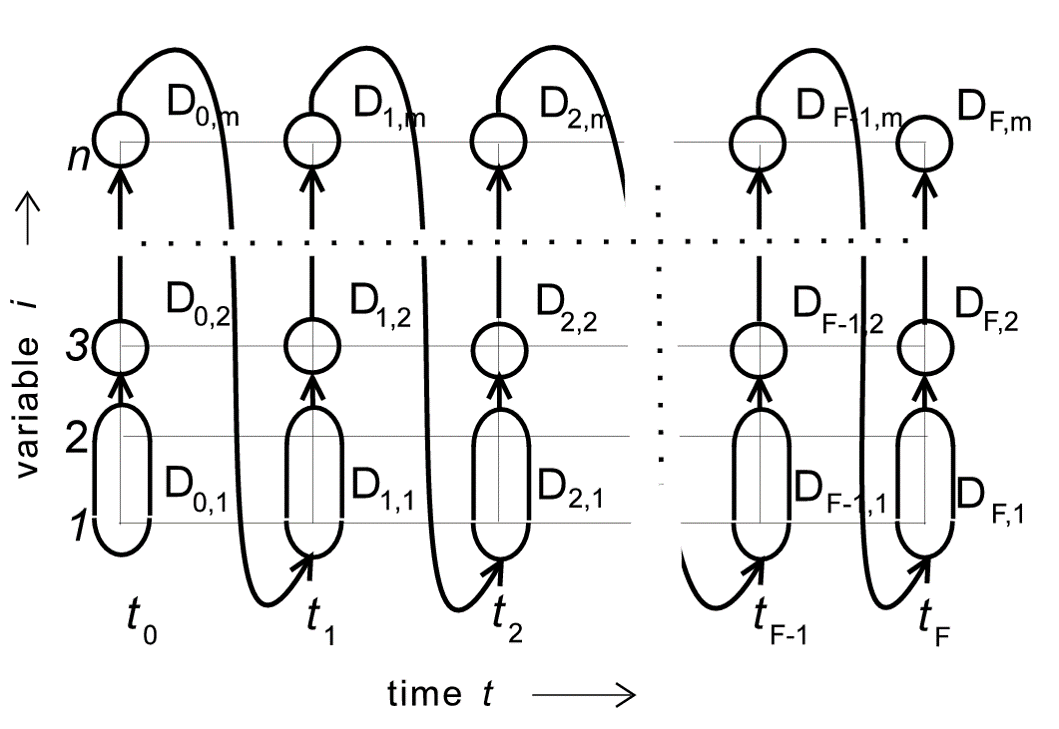}
\caption{The concept of causal relations and the importance of system model \cite{ZampaLecture}. 
\newline There is depicted a set of variables measured at a time instant (represented by circles an ovals) and causal relation in the behaviour between measuring times. To determine the set of measured values at time $ t_{l} $ we must consider not only limited set of measured values at times $ t_{j} < t_{l} $ but also appropriate causality in intermediate times, not defined by the set of system time $ T $. A set of variables at the time of measurement $ D_{k,l} $ is then defined by two indexes $ k,l $, where the first expresses the causality within the measurement time interval, while the other is the index of the system time from the set $ I $. Then, unity $ C_{k,l}\subset \bigcup_{(i,j)<(k,l)}D_{i,j} $ is \textit{the complete immediate cause}, the set of all variables necessary for prediction of the set $ D_{k,l} $. From that, among other conclusions, follows that at least for technical reasons the system may not be understood without the knowledge of an appropriate system model.}
\label{fig:Causal_relations}       
\end{figure*}

Finally, we shall formalise the problem of causality in a measured system (Fig. \ref{fig:Causal_relations}). The system is measured at system times, nevertheless, it is also evolving between them. In the same time, for the definition of the future evolution of the system, it may not be sufficient to consider one time instant, no matter how good our system model is. 

We generally assume that the system trajectory is defined as mapping $z$ with a definition set $D$,

\begin{equation}
D=T\times I,
\label{eq:definition_of_system_trajectory_II}
\end{equation}
which is defined in parts by its internal mechanism. We thus define each segment $z\mid D_{k,l}$ of the trajectory $z$ exactly once and that in the dependency on the segment $z\mid C_{k,l}$, where for $C_{k,l}$ holds

\begin{equation}
C_{k,l}\subset \bigcup_{(i,j)<(k,l)}D_{i,j}.
\label{eq:condition_for_segment_uniqueness}
\end{equation}
Thus, the cause $C_{k,l}$ determines the consequence $D_{k,l}$ and is understood as a \textit{complete immediate cause} of the consequence $D_{k,l}$. 

Thus, we need an appropriate model of the system mainly for two reasons: to determine (1) system behaviour between the time instants and (2) the time extent of the complete immediate cause.

\section{Phenomenological variables and measurement in chemistry}
\label{sec:phenomenological_variables}

The problem of \v{Z}ampa\'{ }s system theory is the definition of a truly appropriate system model. In most cases the models are rather limited, yet, in mechanics or electronics, these limitations may be often overcome. In this discussion we illustrate the problem of measurement in chemistry, which has clear consequences for measurement in biological systems.

In each physico-chemical textbook is introduced the idea of the chemical potential $ \mu_i $ and the activity $ a_i $ which are the real measures of the contribution of each molecule to total Gibbs energy $ G $ of the examined system. It is not from the first sight controversial to write the total Gibbs energy as 

\begin{equation}
G = \sum\limits_{i=0}^n \mu_i,
\label{eq:Gibbs_energy_from_chemical_potentials}
\end{equation}
where index $ i $ determines the individual chemical component of $ n $ components present in the mixture. We should, however, expand $\mu_i$ as  

\begin{equation}
\begin{split}
\mu_i = \mu_{0,i} + \nu_i RT \ln(a_i) = \mu_{0,i} + \nu_i RT\ln ( c_i*\gamma_i ) 
 = \mu_{0,i} + \\ \nu_i RT\ln ( c_i) + \nu_i RT\ln ( f(c_1, c_2... c_n, T, p, V...) ),
\end{split}
\label{eq:activity_definition}
\end{equation}
where $ \mu_{0,i} $ is the standard chemical potential of the i-th component of concentration $ c_i $, $ \gamma_i = f(c_1, c_2... c_n, T, p, V...) $ is its activity coefficient, $ \nu_i $ is the respective stoichiometric coefficient, and $ R $ is the universal gas constant. The activity coefficient is in principle a function of concentrations $ c_i $ of all components in the mixture and all other relevant state variables such as temperature $ T $, pressure $ p $, volume $ V $, etc. The difference between the ideal course, where the $ \gamma_i = 1 $, and a real situation may be demonstrated on such simple examples as distillation of spirits and the existence of the azeotropic mixture. 

From that comes the following moral for the construction of the model of chemical system: we cannot construct our system using concentrations of components as orthogonal variables if we want to obtain a multidimensional plane $ G $  vs. $  \ln  (x_{i}) $. In fact, such a construction is almost never practically possible and the problem is solved by plotting experimental results, giving a complicated surface far from the ideal one. 

The experimentally determined shape of the real state function gives us indices which features we should seek, namely in terms of molecular interactions and their influences on phase behaviour. In the terminology introduced in Chapter \ref{sec:system_theory} we must consider that the definition of system attributes $ A $ and the definition of the appropriate variables $ V $ is inseparable from the system model, i.e., from the mapping $ z $ or the set of probability densities $P$. 

In context of our improper and idealised model, our simply measurable concentrations are \textit{phenomenological variables}. However, in the context of a proper model, describing all molecular interaction and following, e.g., phase changes in macroscopic behaviour, concentrations will be \textit{internal orthogonal variables} of the system. But we completely lose the simple definition of chemical potential $ \mu_{i} $ through logarithm concentration. The relation between chemical potential and concentration may be regained only through the logic of statistical mechanics, specifically through definition of the grand canonic ensemble, and becomes rather impractical. The proper and quite unsatisfying conclusion is that even in chemistry we have the choice between a rather simplified model of low predictive value and the \textbf{phenomenological model} arising from interpolation between experimental variables in the multidimensional space of variables.

\section{Properties of the model of a biological experimental system}
\label{sec:biological_experimental_system}

Biological systems are permanently out of equilibrium state and, moreover, non-homogeneous. In Chapter \ref{sec:phenomenological_variables} was shown that even in the real world of equilibrium chemistry, our idealised models bring us only a very limited insight into the mechanism controlling the system. Nevertheless, they are good communication tools and have been a good start for following, more specific models. In order to get a similar common language, we need adequate discussion tools for biological measurements.  

The most important problem is the long coexistence of several different phases in close distances and its sudden change upon a signal like that for cell division. Biological patterns have been attributed to the periodically repeating solution of the reaction-diffusion equation since the pioneering work of Turing \cite{Turing1952,CrossHohenberg1993}. Similarly, the method of cellular automata has been in parallel developed, i.e. \cite{GreenbergandHastings1978}. In both cases, the final states have been only discussed. These states are not homogeneous in the standard chemical sense, but satisfy the conditions of Poincar\'{e} recurrence \cite{Poincare1890}. Such systems are structured and ergodic, which means that, over sufficiently long periods of time, all accessible microstates are achieved by the system. See, i.e., Birkhoff \cite{Birkhoff1930} for the exact formulation, but one must be aware of the fact that there is a vivid discussion of this problem in physics. 

\begin{figure*}
\centering
\includegraphics[width=0.4\textwidth]{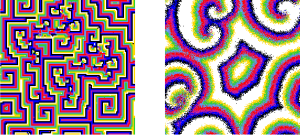}
\caption{Model of the reaction diffusion system using cellular automata \cite{Wuensche2011}. 
\newline The calculation results in an interchanging, dynamic image, which resembles waves, spirals, and related patterns in the Belousov-Zhabotinsky reaction, \textit{Dicyclostelium} colony growth, and other excitable media. The parameters of the models are various rules, threshold interval, etc. The dynamics is sensitive namely to density and type of initial states.}
\label{fig:diffusion_from_DDlam}       
\end{figure*}

There is an obvious physico-chemical problem with these simulations: they ignore many obvious facts, namely coexistence of multiple phases in the living organism. Also, biological systems are \textbf{non-ergodic}. For example, a living cell never visits all possible physical and chemical structures in a time between cell divisions. Quite the contrary -- the cell division seems to be well controlled by various mechanism of its timing (i.e., \cite{Loose2008}). These are two main reasons for insufficiency of contemporary models to provide qualitative basis for methodology of measurement in biological systems.

\begin{figure*}
\centering
\includegraphics[width=0.8\textwidth]{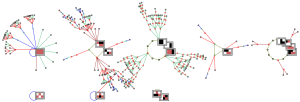}
\caption{Illustration of basins of attraction in a simple discrete system \cite{Wuensche2011}. 
\newline The basin of attraction is a periodically behaving, recurrent, dynamical system, in which structures periodically change or remain stable. In this example, we obtain just 8 non-equivalent basins of attraction. From various initial configurations through variant but defined paths, the system arrives to one of the basins of attraction. In certain cases (in a fixed point), the image remain still. In other cases, there is a periodic interchange between configurations. As seen, paths may be quite long, merge, and diverge.}
\label{fig:DDN}       
\end{figure*}

\begin{figure*}
\centering
\includegraphics[width=0.25\textwidth]{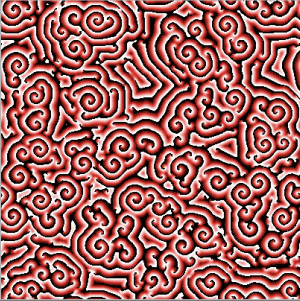}
\caption{Sketch of the basin of attraction of hodgepodge machine (calculated as implemented in the Netlogo software \cite{Wilensky2003}). 
\newline One of the peculiarities of the hodgepodge model is that for its proper functionality it requires only to select an interval of allowed state values (threshold range), or, better expressed, a proper relation between the number of allowed states and the stepwise "chemical reaction" component of the reaction-diffusion simulation. Therefore, it is a good example of a semi-discrete behaviour.}
\label{fig:hodegpodge_analysis}       
\end{figure*}

One of the most visible analogies to the behaviour of biological systems may be found in the domain of cellular automata \cite{Wuensche2011,Shalizietal2002,Crutchfield2011}. The path through the zone of attraction to various attractor basins may be relatively easily searched through correct mathematical simulation of discrete states. Some of these findings have been tracted under the name of discrete dynamic networks by Stuart Kaufmann \cite{Kaufmann1993} (Fig. \ref{fig:DDN}). Perhaps a bit unjustly, in the light of incompleteness of the theory enabling inclusion of the phase transition \cite{Gross2004}, the discrete dynamic networks have been criticized for giving a very little insight into the physico-chemical mechanisms generating living cells. A cellular automaton including three qualitatively different processes -- the Dewndeys hodgepodge model of the Belousov-Zhabotinsky reaction \cite{Dewndey1988} (Fig. \ref{fig:hodegpodge_analysis}) -- is an example. The hodgepodge machine has been much less thoroughly analysed than the Turing pattern, since it is more considered as a "mathematical recreation" than a serious model. 

Since 1990, the recognition of patterns has been extensively studied with the development of machine vision. Each of the recognition methods is based on a certain assumption about the algorithmic principles of generation of the observed image. In the least demanding case of some machine learning approaches, there is an assumption of image statistics. We believe that in this realm we shall seek inspirations for the definition of the proper model of the observed process. 

We have recently contributed to the discussion on general identifiers of observed structures by introduction of a point information gain, point information gain entropy, and point information gain entropy density \cite{Stysetal2011,Stysetal2015}. This approach is based on most general assumption of origins of observed structures both through self-organising processes in the observed object and its projection into the dataset by the measuring device.  

\section{Measurement in biological systems}
\label{sec:measurement_in_biological_systems}

From the reasoning above we may begin to analyse conditions of the design of a proper measurement of biological systems. The main questions are: 
\newline (a) What might be the system attributes -- the set $ A $ -- and what might be variables -- the set $ V $ -- representing them? 
\newline (b) What is the proper model of the system? 

We may start with the problem of discreteness. Usage of discrete entities such as agents or pixels has provided many good analogies to observations in biological and social systems. Unfortunately, the nature is not exactly discrete but only partly discrete, i.e., each animal is composed of -- distinct -- organs, organs of -- distinct -- cells etc.  And these organs, cells etc. are again existent in discrete states. The immediate objection to the previous statements is that organs or cells are not as distinct as elementary pixels in the computer simulation and their states can not be described with natural numbers. However, to some extent, the biological experience does that and the biological literature is full of precise statements on cell, organ, organism, etc. states. 

Similarly, as we are unable to measure at the infinite number of the time instants, we usually measure only a sub-set of possible variables. If only one of the \textbf{system trajectories} or, even worse, one of the cell states is selected and reported, a minute sub-section of the whole problem is obviously obtained. This is neither a new nor surprising problem -- in Feymann's worlds "you should always decide to publish it whichever way it comes out" \cite{Feymann1973}. But in biological systems the problem is more serious, as it is complicated by the strong non-linearity which is selecting the basin of attraction by initial conditions. Thus, the scope of results is limited, sometimes in a highly reproducible manner. The problem of irreproducibility of biological results arises from "not publishing whichever comes out" further amplified by the fact that a very constrained sub-set of outcomes is obtained at given time, with given strain and set of chemicals. But that is only due to a very specific set-up when many conditions are not recorded.

In Chapter \ref{sec:phenomenological_variables} we have discussed the problem of chemical activity vs. concentration with the conclusion that with the usage of an adequate model, concentration may be a good orthogonal coordinate of the system model. Just, the surface describing the multidimensional state equation $ G = f( c_1, c_2, ... c_n, p, V, T,... ) $ might be quite complicated. In the case of systems highly sensitive to initial conditions we might expect an occurrence of highly divergent trajectories originating from very well determined initial conditions. And these trajectories themselves might be confined to rather small region of the phase space. In this case, we may define true phenomenological variables as a set of variables leading to the same state trajectory. In other words, the set of trajectories determining the system event $ B $ is not arbitrary, but determined by system internal behaviour -- which may be also understood as the best possible model. We propose to name such a distinct system event as \textit{a system phenomenon} $ \Phi $, where

\begin{equation}
\Phi \subset \Omega,
\label{eq:definition_of_system_phenomenon}
\end{equation}
and propose that signs of the system phenomena should be defined and examined. As we explained before, there is a good, theoretically substantiated, reason that even a quite small subset of the measured system variables will give a good stochastic model. Based on this good discrimination, we may define a decomposition $ F $  of the set $ \Omega $ into disjunct subsets $ \Phi $: 

\begin{equation}
\Phi \subset F.
\label{eq:definition_of_system_phenomenonII}
\end{equation}

The stochastic system $ {\mathscr{S}}=(T,V,P) $ may be replaced by the phenomenological stochastic system $ {\mathscr{F}}=(T,V,P_{\phi}) $, where $ P_{\phi} $ is the probability of transition between individual phenomena at given combination $ (T,V) $. 

In case of knowledge of an appropriate non-linear model, it is enough to know the resulting trajectory at the given set of conditions. Or, in other words, we can examine the position of the border between the two zones of attraction experimentally, instead of common "constructive" examination of conditions, i.e., variable values, at which the system works in a desired way. 

The state of the organism is known only if a complete phenomenological model $ \mathscr{F} $ is known. This is in most cases practically impossible. It may be said with certain exaggeration that the only statistically relevant biological experiment is a record of the distribution of stock exchange indexes. Thus, we propose a different framework for reporting and analysis of biological experiments. The most precise possible description of each experimental step is a must. Under any circumstances we must anticipate that the biological object may follow a wildly different trajectory due to subtle differences in the set-up, which were not reported. The behaviour of biological systems has to be studied in the framework of mathematics of non-linear systems outside the Lyapunov stability. And this is so far almost unstudied problem, namely since it is not understood as such. 

\begin{figure*}
\centering

\begin{tabular}{c c c}

\includegraphics[width=2.7cm]{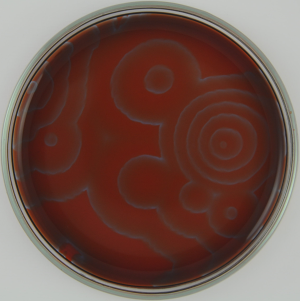}&
\includegraphics[width=2.7cm]{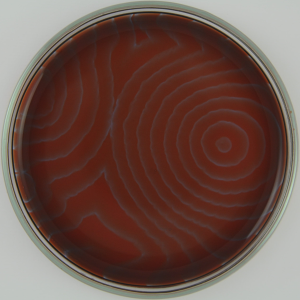}&
\includegraphics[width=2.7cm]{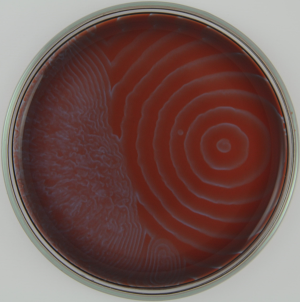}
\\
\includegraphics[width=2.7cm]{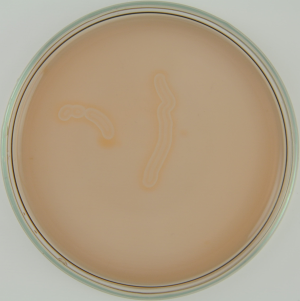}&
\includegraphics[width=2.7cm]{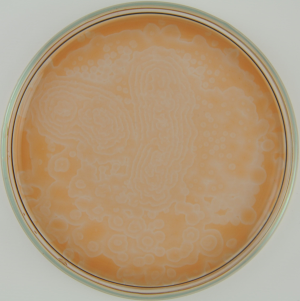}&
\includegraphics[width=2.7cm]{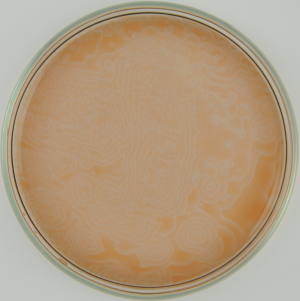}

\end{tabular}
\caption{The course of the Belousov-Zhabotinsky reaction as a consequence of different purity of a reactant. 
\newline The experiments were performed as described in the commercially available kit \cite{Cohen2010} with the difference in a supplier of ferroin. $Upper$  $row$ -- supplier Penta (the time instants indexes $ i $ were 25, 50, and 100), $lower$ $row$ -- supplier Fluka  ($i$ = 5, 15, and 30). Distance between time instants was 10 s.}
\label{fig:BZ_using_different_chemicals}       
\end{figure*}

Instead of showing a biological example -- which would be difficult to explain in a sketchy way -- we demonstrate our idea using a much simpler example of chemical self-organisation, the Belousov-Zhabotinsky reaction. In Fig. \ref{fig:BZ_using_different_chemicals} we show the performance of this well known experiment using chemicals obtained from two different providers. Although the ferroin indicator was in both cases declared to be of p.a. quality,  we have obtained two wildly different self-organising structures. To give an illustrative explanation let us consider following: if the presence of the contamination in the chemical is guaranteed to be less than 0.1\% , there is still $ 10^{19} $ of undetected molecules in each mole of the chemical. In case of sensitivity to initial conditions, the difference in the behaviour is not surprising. 

\section{Conclusions}
\label{sec:conclusions}

There are two main domains in any experiment -- the design of the experiment and analysis of its results. In this article, we show that the usage of tools for design of experiment, which have been well established in physical and many chemical experiments is no guarantee for the proper experiment design in the biological (and many chemical) experiments. 

Proper discussion of this discrepancy should be made in the frame of the theory of dynamic systems (Chapter \ref{sec:system_theory}). The cybernetic systems analysis anticipates that each of the experiments may be performed from the beginning, in another words that the instant $ t_0 $ of the experimental time is equal to $ t=0 $ of the time of the system dynamics -- e.g., on/off change of the state of an electrical switch. Or, more exactly, we assume that all time instants included in the complete immediate cause $C_{k,l}$ determining the behaviour of the system at time of the measurement $ t_k $ lie within the time extent of the set $ T $. Also, in a standard experiment, we consider that we are able to determine values of sufficient number of variables from the set $ V $, which allow us to build good model of the experiment. 

In equilibrium physical chemistry, we rather assume the occurrence of the system on the manifold of the state equation. This also assumes that for actual values of state variables any history of the system is irrelevant, i.e., $C_{k,l}=D_{k,l}$. 

In biology, we should instead a-priori assume that the complete immediate cause $C_{k,l}$ may not only contain all time instants covered by the measurement, but that it may even include some events which occurred at time instants before the experiment has started. 

The contemporary unspoken common assumption is that biological system may by described as a special chemical system. Since as early as 17$^{th}$ century, some thinkers have been assuming that biological systems may be modelled as (in present terminology) cybernetic machines. The possibility that the biological system may be understood as an equilibrium chemical system is clearly incorrect. But also the fact that a biological system may be, to some extent, modelled as a cybernetic system comes from the biased interpretation of its non-linearity and semi-discreteness. It is for that reason that we observe only few basins of attraction and a few paths through the state space which lead to them. Finally, the path, within which the phenomenological stochastic system $ {\mathscr{F}}=(T,V,P_{\phi}) $ is evolving in time, may include a much smaller part of the whole phase space than that in which a mechanical system with the Gaussian probability density function is evolving. This may be mistaken with a good reproducibility of the biological experiment when it is repeated with the same set of chemicals and within a short time interval of repetition. It leads on one hand to relative sloppiness in the definition of experimental conditions and on the other hand to bad surprises, when the experiment needs to be reproduced or transferred to production line \cite{Prinz2011,Begley2012}. 

The proper conclusion is that a biological experiment will never be complete, simply because we can never reverse the time and we shall not know the true starting point. One of the possibilities how to proceed in a proper analysis of the biological experiment is to seek conditions at which the system begins to follow another trajectory. It is similar to the qualitative analysis of the system of non-linear differential equations when nullclines are sought \cite{KlippetalSystemsBiology}. However, we must be aware of the fact that our system is in-part discrete, which means that we rather seek trajectories to the basin of attraction in a cellular automaton as shown by Wuensche \cite{Wuensche2011}. This factor of semi-discretion leads to the positive role of noise in biology \cite{Tsimrig2014}, which we may briefly describe as a constant faltering of the system, which may more frequently occupy trajectory acquiring a broader part of the phase space. 

The biological system is also periodically internally re-started and\linebreak re-synchronised. Let us mention the control of the bacterial cell cycle by\linebreak MinD/MinE system \cite{Loose2008} as an illustrative example. We have recently shown \cite{Stysetal2015Purplsoc} that the method of shaking influences strongly the outcome of the self-organisation in the Belousov-Zhabotinsky reaction. 

An obvious solution to the problem of measurement in biological systems is to record as many experimental outcomes as possible and publish them. This does not satisfy the human desire of understanding the system, i.e., making a model for the given observation. Yet, we may gradually come close to the suitable phenomenological description of the relatively narrow distribution of the distinct possible outcomes. This possibility comes from strong non-linearity and results in tendency to classify biological phenomena qualitatively, i.e., giving it a name such as "stress behaviour", "resting state", etc. These are the phenomena $ \Phi $ discussed in Chapter \ref{sec:measurement_in_biological_systems}. The persistent problem is how to define them properly. In our opinion, many jewels are hidden in \textbf{experimental protocols and their evolution}, whose analysis may lead to the proper classification and construction of a really suitable model. Experimental protocols often evolve from a simple set-up of chemical type into elaborate knowledge including provider of chemicals and many tricks, often unspoken. But only such analysis eventually leads to a successful biotechnological procedure or a relatively reproducible experiment.   

Our knowledge-based data repository bioWes \cite{Biowesweb} provides solution to the problem. Its key component is the protocol generator which records the evolution of protocols. To each individual protocol is attached the respective dataset. We believe that the bioWes approach may lead to true understanding of biological systems as well as to, e.g., acceleration of development and increase of reliability of biotechnological drugs. 

\section{Acknowledgement}

This work was partly supported by the Ministry of Education, Youth and Sports of the Czech Republic -- projects CENAKVA (No. CZ.1.05/2.1.00/01.0024) and CENAKVA II (No. LO1205 under the NPU I program), by Postdok JU \linebreak CZ.1.07/2.3.00/30.0006, and GAJU Grant (134/2013/Z 2014 FUUP). Authors thank to Petr Jizba, Jaroslav Hlinka, Harald Martens, \v{S}t\v{e}p\'{a}n Pap\'{a}\v{c}ek and Tom\'{a}\v{s} N\'{a}hl\'{i}k for important discussions.


%
\end{document}